\documentclass[aps,amsmath,amssymb]{revtex4}
\usepackage{amsfonts}
\usepackage{amssymb}
\usepackage{graphicx}
\usepackage{subfigure}
\usepackage{color}
\usepackage[normalem]{ulem}
\usepackage{bbold}
\usepackage{placeins}
\usepackage{soul,xcolor}
\usepackage{epstopdf}
\usepackage{tabu}
\usepackage{array}
\usepackage{upgreek}
\usepackage{multirow}
\usepackage[utf8]{inputenc}
\usepackage[colorlinks = truelinkcolor = blue, urlcolor  = blue, citecolor = blue, anchorcolor = blue]{hyperref}
\hypersetup{
	colorlinks   = true, %Colours links instead of ugly boxes
	urlcolor     = blue, %Colour for external hyperlinks
	linkcolor    = blue, %Colour of internal links
	citecolor   = blue %Colour of citations
}

\newcommand{\be}{\begin{equation}}
\newcommand{\ee}{\end{equation}}
\newcommand{\ba}{\begin{array}}
\newcommand{\ea}{\end{array}}
\newcommand{\bea}{\begin{eqnarray}}
\newcommand{\eea}{\end{eqnarray}}

\def\bra#1{\left\langle #1\right|}
\def\ket#1{\left| #1\right\rangle}

\begin{document}

	\title{Study of temporal quantum correlations in decohering $B$ and $K$ meson systems}
	
    \author{Javid Naikoo}
	\email{naikoo.1@iitj.ac.in}
	\affiliation{Indian Institute of Technology Jodhpur, Jodhpur 342011, India}
	
	\author{Ashutosh Kumar Alok}
	\email{akalok@iitj.ac.in}
	\affiliation{Indian Institute of Technology Jodhpur, Jodhpur 342011, India}
	
	\author{Subhashish Banerjee}
	\email{subhashish@iitj.ac.in}
	\affiliation{Indian Institute of Technology Jodhpur, Jodhpur 342011, India}

\date{\today} 
	
\begin{abstract}
In this work we study temporal quantum correlations, quantified by Leggett-Garg  (LG) and LG-type inequalities, in the $B$ and $K$ meson systems. We use the tools of open quantum systems to incorporate the effect of decoherence which is quantified by a single phenomenological parameter. The effect of $CP$ violation is also included in our analysis.  We find that the LG inequality is violated for both $B$ and $K$ meson systems, the violation being most prominent in the case of $K$ mesons and least for $B_s$ system.  Since the systems with no coherence do not violate LGI, incorporating  \textit{decoherence} is expected to decrease the extent of violation of LGI and is clearly brought out in our results.  We show that the  expression for the LG functions depends upon an additional term, apart from the experimentally measurable meson transition probabilities. This term  vanishes in the limit of zero decoherence.  On the other hand, the LG-type parameter can be directly expressed in terms of transition probabilities, making it a more appropriate observable for studying temporal quantum correlations in neutral meson systems. 

\end{abstract}

\maketitle 
	
%%%%%%%%%%%%%%%%%%%%%%%
\section{INTRODUCTION} 
\label{intro}
%%%%%%%%%%%%%%%%%%%%%%

 Quantum correlations (QCs), existing between two or more parties \cite{horodecki2009quantum, modi2012classical}, are bestowed with properties unique to the quantum world and are of pivotal importance in quantum information science. The study of QCs not only unveils the fundamental traits responsible for the distinction of the quantum mechanically correlated systems from those attributed with a joint classical probability distribution \cite{BellCHSH}, it also helps in devising efficient ways of carrying out the tasks of quantum communication and computation \cite{gisin2002quantum, raussendorf2001one, briegel2009measurement}.  
 
    Among the most celebrated notions in quantum physics are nonlocality \cite{bell1964einstein}, entanglement \cite{horodecki2009quantum}, quantum discord \cite{ollivier2001quantum,hendersonvedral}, and teleportation fidelity \cite{barrett2004deterministic, sbdistel}. These spatial quantum correlations (SQCs) have enhanced our understanding of nature at the fundamental level and at the same time have provided efficacious solutions in the development of the theory of quantum information. The SQCs mentioned above have been  studied in many systems, viz., optical systems \cite{aspect1981experimental, tittel1998experimental, tittel1998violation, weihs1998violation, indranilsb,lanyon2013experimental, naikoo2017probing}, NMR \cite{ kessel2000quantum, dorai2001quantum, laflamme2001nmr}, neutrino oscillation \cite{blasone2009entanglement, alok2016quantum, banerjee2015quantum, Naikoo:2017fos}, $B$ and $K$ meson systems \cite{banerjee2016quantum}. Of the above listed SQCs, Bell nonlocality is the strongest and Bell inequalities are considered to be the oldest tool for detecting entanglement  \cite{guhne2009entanglement}.
    
    The temporal quantum correlations (TQCs) arising from the sequential measurements on a system at different times, have also been considered as promising candidates in discerning the quantum behavior from the classical. Leggett and Garg inequalities (LGIs) \cite{leggett1985quantum} are among the well known TQCs, violation of which is a witness of quantum \textit{coherence} in the system. LGIs have been a topic of study in various theoretical works  \cite{barbieri2009multiple, avis2010leggett, lambert2010distinguishing, lambert2011macrorealism, montina2012dynamics, emary2013leggett, kofler2013condition} including, in recent times, neutrino oscillations \cite{formaggio2016violation,  Naikoo:2017fos, Fu:2017hky}  and studied experimentally in  systems like superconducting qubits \cite{palacios2010experimental, groen2013partial}, photons \cite{goggin2011violation, xu2011experimental, dressel2011experimental, suzuki2012violation},  and NMR \cite{athalye2011investigation, souza2011scattering, katiyar2013violation}.

     Leggett-Garg inequalities are based on the concept of \textit{macrorealism} (MR) and \textit{noninvasive measurability} (NIM). MR means that the system which has available to it two or more macroscopically distinct states, pertaining to an observable $\hat{Q}$, always exists in one of these states irrespective of any measurement performed on it. NIM states that, in principle, we can perform the measurement without disturbing the future dynamics of the system \cite{emary2013leggett}.  MR and NIM put limits on  certain combinations of the two time correlation functions $C_{ij} = \langle \hat{Q(t_i)} \hat{Q(t_j)} \rangle $. Quantum systems, however, violate these limits. The simplest form of LGI is the one involving three measurements performed at $t_0$, $t_1$ and $t_2$ ($t_2 > t_1 > t_0$)
    \begin{equation}\label{K3defined}
    K_3 = C_{01} + C_{12} - C_{02},
    \end{equation}
    such that $-3 \le K_{3} \le 1$. The maximum quantum value of $K_3$ for a two level system in $\frac{3}{2}$ \cite{leggett1985quantum} and has been found to hold for any system, \textit{irrespective} of the number of levels, as long as the measurements are given by just two projectors $\Pi^{\pm}$ \cite{budroni2013bounding}, a fact revealed in several studies \cite{george2013opening, kofler2007classical, lambert2011macrorealism, wilde2010could}. It was shown in \cite{budroni2014temporal} that in the limit $N \rightarrow \infty$, the LGI can be violated up to its maximum algebraic sum. 
 
 The autocorrelation $C_{12}$ turns out to contain a nonmeasurable quantity and hence reduces the efficacy of Eq. (\ref{K3defined}) from the experimental point of view. Such limitations of the two time correlations have been discussed in detail in \cite{huelga1995proposed, huelga1996temporal, huelga1997observation, waldherr2011violation}, and a different approach was developed which involves replacing the NIM by a weaker condition called  \textit{stationarity}. This avoids the need of performing the measurement at the intermediate time $t_1$ by replacing $C_{12}$ by $C_{01}$, thus leading to an easily testable Leggtt-Garg type inequality (LGtI)   
    \begin{equation}\label{tildeK3defined}
   \tilde{ K}_3 = 2C_{01} - C_{02} \le 1.
    \end{equation} 
The following set of assumptions  \cite{emary2013leggett} are considered important for applying \textit{stationarity} to a system: (i) macroscopic realism, (ii) the conditional probability $P(\psi, t+t_0| \psi, t_0)$ of finding the system in state $\psi$ at time $t+t_0$ given that it was in state $\psi$ at time $t_0$, should be invariant  under the time translation, $P(\psi, t+t_0|\psi,t_0) = P(\psi,t|\psi,0)$, (iii)  Markovianity and (iv) that the system is prepared in state $\psi$ at time $t=0$.

 In this work, we study the LG and LG-type inequalities in the $B$ and $K$ meson systems. The effect of decoherence is included by using the formalism of open quantum systems. Decoherence, here, is modelled by a single phenomenological parameter \cite{Alok:2015iua} which represents the interaction between the one-particle system and its environment. The environment can be attributed to quantum gravity effects \cite{qg1,qg2,qg3,qg4,qg5,qg6,qg7,qg8} or it can be due to detector background itself.
Apart from decoherence, we also include the effects of $CP$ violation. We find that the LG inequality is violated for both $B$ and $K$ meson systems. Apart from the experimentally measurable meson transition probabilities,  we show that the LG function depends upon an additional term which vanishes in the limit of zero decoherence.  The LG-type parameter  on the other hand can be directly expressed in terms of transition probabilities.

The plan of this work is as follows. In the next section, we discuss the time evolution of $B$ and $K$   meson systems treated as open quantum systems. In Sec. III, we derive the LG and LG-type inequalities for these systems. In Sec. IV, we present our results. Finally, in Sec. V, we make our conclusions.

%%%%%%%%%%%%%%%%%%%%%%%%%%%%%%%%%%%%
	\section{$B$ and $K$ mesons as open quantum systems}\label{dynamics}
%%%%%%%%%%%%%%%%%%%%%%%%%%%%%%%%%%%%%]

 In this section, we introduce our formalism for the study of $B$ and $K$ mesons as open quantum systems.

	 \subsection{Kraus representation} 
    Kraus representation \cite{kraus1983states}, describes the time evolution of an \textit{open} quantum system, which is not necessarily unitary unlike the evolution of a \textit{closed} quantum system. Real physical systems are always entangled with their ambient environment, alternatingly addressed as the reservoir. Kraus representations are very convenient for handling a number of practical problems of open system dynamics \cite{ breuer2002theory, weiss2012quantum, nielsen2000quantum, banerjee2010entanglement, banerjee2010dynamics, omkar2012operator}. 	Consider a large system $S$ comprising of two subsystems $S_a$ and $S_b$. At a given time $t$, let the quantum states corresponding to $S$, $S_a$ and $S_b$ be represented by $\rho(t)$, $\rho_a(t)$ and $\rho_b(t)$, respectively. Then $\rho_a(t) = Tr_b \{\rho(t) \}$ and $\rho_b(t) = Tr_a \{\rho(t) \}$. Since the total system is unitary, its evolution is given by 
    \begin{equation}
    \rho(t) = U(t) \rho(0) U^{\dagger}(t), 
    \end{equation}
    where $U(t)$ is a unitary operator.The evolution of system $S_a$ will look like
    \begin{equation}
    \rho_a(t) = Tr_b\{U(t)\rho(0) U^{\dagger}(t)\}. \label{rho_a}
    \end{equation}
	If it is possible to recast Eq. (\ref{rho_a}) in the following form
	\begin{equation}
	\rho_a(t) = \sum_{i} E_i(t) \rho_a(0) E_i^{\dagger}(t),
	\end{equation}
	such that $\sum_{i} E_i(t) E_i^{\dagger}(t) = \mathbb{1} $, then the evolution of  $\rho_a(t)$  has a $Kraus$ representation and is completely positive.

\subsection{Time evolution of $B$/$K$ mesons}	
We describe briefly the time evolution of $B^o(K^o)$ meson system. Since both $B^o$ and $K^o$ share the same scheme of dynamics, we discuss only $B^o$ system and the results, with appropriate notational changes, will be applicable to the $K^o$ system. The states of the total system, including the meson and the vacuum $\ket{0}$, introduced in order to incorporate the effect of decay in the meson system, reside in the Hilbert space given by the direct sum $\mathcal{H}_{B^0}  \oplus \mathcal{H}_{0}$ \cite{caban2005unstable, Alok:2015iua, Alok:2013sca} spanned by the orthonormal vectors $\ket{B^0}$, $\ket{\bar{B}^0}$ and $\ket{0}$
 
 \begin{equation}
 \ket{B^0} = \begin{pmatrix}
                     1  \\
                     0   \\
                     0
             \end{pmatrix}; \quad \ket{\bar{B}^0} = \begin{pmatrix}
                                                       0  \\
                                                       1   \\
                                                       0 
                                                    \end{pmatrix}; \quad  \ket{0} = \begin{pmatrix}
                                                                                        0  \\
                                                                                        0   \\
                                                                                        1
                                                                                      \end{pmatrix}. \label{basis}
 \end{equation} 
Here $B^0$ stands for $B^0_d/B^0_s$ mesons. The mass eigenstates $\{\ket{B_L}, \ket{B_H} \}$ are related to the flavor eigenstates $\{ \ket{B^o}, \ket{\bar{B}^o}\}$ by the equations 
\begin{equation}\label{MassFlavorStates}
\ket{B_L} = p \ket{B^o} + q \ket{\bar{B}^o}, \qquad \ket{B_H} = p \ket{B^o} - q \ket{\bar{B}^o},
\end{equation} 
	with $|p|^2 + |q|^2 = 1$.	The time evolution is given by a family of completely positive trace preserving maps forming a one parameter dynamical semigroup. The complete positivity requires the time evolution of a state of the system being represented by the operator-sum representation  \cite{kraus1983states}

\begin{equation}
\rho(t) = \sum_{i=0} E_{i}(t) \rho(0) E^{\dagger}_{i}(t), \label{operator_sum_rep}
\end{equation}
where the $Kraus$ operators have the following form

 \begin{align*}
E_0 &= \ket{0}\bra{0},\\
E_1 &= \mathcal{E}_{1+}\big( \ket{B^0}\bra{B^0} + \ket{\bar{B}^0}\bra{\bar{B}^0}\big ) + \mathcal{E}_{1-}\big( \frac{p}{q}\ket{B^0}\bra{\bar{B}^0} + \frac{q}{p}\ket{\bar{B}^0}\bra{B^0} \big),\\
E_2 &= \mathcal{E}_2 \big( \frac{p+q}{2p} \ket{0}\bra{B^0} + \frac{p+q}{2q} \ket{0}\bra{\bar{B}^0} \big), \\
E_3 &= \mathcal{E}_{3+} \frac{p+q}{2p} \ket{0}\bra{B^0} + \mathcal{E}_{3-} \frac{p+q}{2q} \ket{0}\bra{\bar{B}^0} , \\
E_4 &= \mathcal{E}_4 \big( \ket{B^0}\bra{B^0} + \ket{\bar{B}^0}\bra{\bar{B}^0} + \frac{p}{q}\ket{B^0}\bra{\bar{B}^o} + \frac{q}{p}\ket{\bar{B}^0}\bra{B^0} \big),\\
E_5 &= \mathcal{E}_5 \big( \ket{B^0}\bra{B^0} + \ket{\bar{B}^0}\bra{\bar{B}^0} - \frac{p}{q}\ket{B^0}\bra{\bar{B}^0} - \frac{q}{p}\ket{\bar{B}^0}\bra{B^0} \big).
\end{align*}
Here the coefficients are 
\begin{widetext}
	\begin{subequations}
	\begin{align}
\mathcal{E}_{1\pm} &= \frac{1}{2} \left[ e^{-(2 i m_L + \Gamma_L + \lambda) t/2} \pm e^{-(2 i m_H + \Gamma_H + \lambda) t/2} \right], \label{E1}\\
\mathcal{E}_2  &= \sqrt{\frac{Re[\frac{p-q}{p+q}]}{|p|^2 - |q|^2} \big( 1 - e^{-  		\Gamma_L t} - (|p|^2 - |q|^2)^2  \frac{|1 - e^{-(\Gamma + \lambda - i \Delta m )t}|^2}{1 - e^{-\Gamma_H t}}}\big), \label{E2} \\
\mathcal{E}_{3\pm} &= \sqrt{\frac{Re[\frac{p-q}{p+q}]}{(|p|^2 - |q|^2)(1 - e^{-\Gamma_H t})}}\big[1 - e^{-\Gamma_H t}  \pm (1 - e^{-(\Gamma + \lambda - i \Delta m)t})(|p|^2 - |q|^2)\big], \label{E3}\\
\mathcal{E}_{4} &= \frac{e^{-\Gamma_L t/2}}{2} \sqrt{1 - e^{-\lambda t}},\label{E4}\\
\mathcal{E}_{5} &= \frac{e^{-\Gamma_H t/2}}{2} \sqrt{1 - e^{-\lambda t}}. \label{E5}
   \end{align}
   \end{subequations}
 
 A meson initially in state $\rho_{B^0}(0) = \ket{B^0}\bra{B^0}$ or $\rho_{\bar{B}^0}(0) = \ket{\bar{B}^0}\bra{\bar{B}^0}$, after time $t$, evolves to 

 \begin{equation}
 \rho_{B^0}(t) = \frac{1}{2}e^{-\Gamma t} \begin{pmatrix}
                                          a_{ch} + e^{-\lambda t} a_{c}                   & (\frac{q}{p})^* (-a_{sh} - i e^{- \lambda t} a_s)     &       0 \\
                                          (\frac{q}{p}) (-a_{sh} + i e^{-\lambda t} a_s)  & |\frac{q}{p}|^2 a_{ch} - e^{-\lambda t} a_{c}         &       0  \\
                                                  0                                       &                0                                      &     \rho_{33}(t)
                                    \end{pmatrix} \label{rhoBt},
 \end{equation}
 and 
 \begin{equation}
 \rho_{\bar{B}^0}(t) = \frac{1}{2} e^{-\Gamma t} \begin{pmatrix}
                                                  |\frac{p}{q}|^2 (a_{ch} - e^{- \lambda t} a_c)      &   (\frac{p}{q}) (-a_{sh} + i e^{- \lambda t} a_{s})  &  0  \\
                                                  (\frac{p}{q})^* (-a_{sh} -i e^{- \lambda t} a_{s})  &    a_{ch} + e^{- \lambda t} a_c                      &  0   \\
                                                        0                                             &           0                                          &  \tilde{\rho}_{33}(t) \\
                                                 \end{pmatrix}  \label{rhoBbart}.
 \end{equation}
\end{widetext}
 Here, $a_{ch}$ ( $a_{sh}$)  and $a_{c}$ ($a_{s}$) stand for the hyperbolic functions $\cosh[{\frac{\Delta \Gamma t}{2}]}$ ($\sinh{[\frac{\Delta \Gamma t}{2}]}$) and the trigonometric functions $\cos{[\Delta m t]}$ ($\sin{[\Delta m t]}$), respectively. $p$ and $q$ are defined in Eq. (\ref{MassFlavorStates}). $\Delta\Gamma = \Gamma_L - \Gamma_H$ is the difference of the decay width $\Gamma_L$ (for $B^o_L$ ) and  $\Gamma_H$ (for $B^o_H$). $\Gamma = \frac{1}{2}(\Gamma_L + \Gamma_H)$ is the average decay width. The mass difference $\Delta m = m_H - m_L$, where $m_H$ and $m_L$ are the masses of $B^o_H$ and $B^o_L$ states, respectively. The strength of the interaction between the one particle system and its environment is quantified by $\lambda$, the \textit{decoherence} parameter \cite{ABUDecho}. The elements $\rho_{33}(t)$ and $\tilde{\rho}_{33}(t)$ are known functions of B physics parameters, not used in this work. In the following section, we use this formalism to develop the LGI and LGtI for the meson systems.

%%%%%%%%%%%%%%%%%%%%%%%%%%%%%%%%%%%%%%%%%%%%%%%%%%%%%%%%%
\section{Temporal quantum correlations in B/K systems}\label{LGI_for_BKmesons}
%%%%%%%%%%%%%%%%%%%%%%%%%%%%%%%%%%%%%%%%%%%%%%%%%%%%%%%%%

\subsection{Leggett-Garg inequality}

Leggett-Garg inequalities, often referred to as the temporal Bell inequalities, place bounds on certain combinations of the two time autocorrelations $C_{ij}$, defined in terms of the joint probabilities as \cite{leggett1985quantum, kofler2008conditions, castillo2013enhanced}
	     \begin{equation}
	     C_{ij} = p(^{+}t_i)q(^{+}t_j|^{+}t_i) - p(^{+}t_i)q(^{-}t_j|^{+}t_i) - p(^{-}t_i)q(^{+}t_j|^{-}t_i) + p(^{-}t_i)q(^{-}t_j|^{-}t_i) \label{Cij_pq},
	     \end{equation}
	     where $p(^{a}t_i)$ is the probability of obtaining the  result $a = \pm1$ at $t_i$, and $q(^{b}t_j|^{a}t_i)$ is the conditional probability of getting result $b=\pm 1$ at time $t_j$, given that result $a = \pm 1$ was obtained at $t_i$. To find the probabilities involved in Eq. (\ref{Cij_pq}), we define the projector $\Pi^{\pm}$ related to the eigenspace  of the dichotomic operator $\hat{Q}$, such that the probability of obtaining outcome $a$ at time $t_i$ is 
	     \begin{equation}
	     p(^{a}t_i) =  Tr\{\Pi^a \rho(t_i)\} = Tr\{\Pi^a \sum_\mu K_\mu(t_i) \rho(0) K^{\dagger}_\mu(t_i) \}.
	     \end{equation}
	     The density matrix corresponding to the measurement result $a$ obtained at `$t_i$' is given by the von Neumann rule
	     \begin{equation}
	     \rho^a(t_i) = \frac{\Pi^a \rho(t_i) \Pi^a}{Tr\{\Pi^a \rho(t_i)\}} = \frac{\Pi^a \sum_\mu K_\mu(t_i) \rho(0) K^{\dagger}_\mu (t_i) \Pi^a}{p(^at_i)},
	     \end{equation} 
	     this state evolves until $t_j$, when the state of the system looks like $\sum_\nu K_\nu(t_j - t_i) \rho^a(t_i) K^{\dagger}_\nu(t_j - t_i)$, so that the probability of obtaining outcome $b$ at time $t_j$, given that $a$ was obtained at time $t_i$, is given by
	     
	     \begin{equation}
	     q(^bt_j|^at_i) =  \frac{Tr\{\Pi^b \sum_{\nu, \mu} K_\nu(t_j - t_i) \Pi^a K_\mu(t_i) \rho(0) K^{\dagger}_\mu (t_i) \Pi^a K^{\dagger}_\nu(t_j - t_i)\}}{p(^at_i)}.
	     \end{equation} 
	
	     A generic term in the right-hand side of Eq. (\ref{Cij_pq}) becomes 
	     \begin{equation}
	     p(^{a}t_i)q(^{b}t_j|^{a}t_i) = Tr\{\Pi^b \sum_{\nu} K_{\nu,\mu}(t_j - t_i) \Pi^a K_{\mu} (t_i) \rho(0) K_{\mu}^{\dagger}(t_i) \Pi^{a} K^{\dagger}_{\nu}(t_j - t_i)\}. \label{generic_term}
	     \end{equation} 
	     With some algebra, we can show that the two time correlations turn out to be \cite{castillo2013enhanced}
	     \begin{equation}
	     C_{ij} = 1 - 2p(^+t_1) - 2p(^+t_2) + 4Re[g(t_i, t_j)],
	     \end{equation}	   
	     where
	     \begin{equation}
	     g(t_i,t_j) = Tr\Big\{\Pi^+ \sum_{\nu} K_\nu(t_j - t_i) \Pi^+ \rho(t_i) K^{\dagger}_\nu(t_j - t_i)\Big\}.
	     \end{equation}

 We consider a dichotomic quantity $Q=\pm1$ for our \textit{three} level system, such that each level is associated with a definite value of Q. Assigning the same value of Q to different states is irrelevant from the macrorealistic point of view and does not change the bounds of Eq. (\ref{K3defined}) \cite{budroni2014temporal}. Let us assume that at time $t=0$, the meson was in state $\rho_{\bar{B}^0}$. This state evolves to $\rho_{\bar{B}^0(t_i)}$ at time $t_i$ and is given by Eq. (\ref{rhoBbart}). We define the dichotomic operator $\Pi = \Pi^+ - \Pi^- = \Pi_{B^0} - (\Pi_{\bar{B}^o} + \Pi_{0})$, where $\Pi_x = \ket{x}\bra{x}$. Now
  \begin{equation}
p(^+t_i) = Tr\{\Pi^+ \rho_{\bar{B}_0}(t_i)\} =  [\rho_{\bar{B}_0}(t_i)]_{11}
         = |p/q|^2 \frac{e^{-\Gamma t_i}}{2} \bigg[ \cosh(\frac{\Delta \Gamma t_i}{2}) - e^{-\lambda t_i} \cos(\Delta m t_i)\bigg] \label{Survival}.
  \end{equation}
Thus, $p(^+t_i) = \mathcal{P}_{\bar{B^0} B^0}(t_i)$ is the transition probability from state $\rho_{\bar{B}^0}$ to $\rho_{B^0}$ at time $t_i$. With the assumption of equal time measurements $t_2-t_1 = t_1 - 0 = \Delta t$, we have the following expression for $C_{12}$
\begin{equation}
C_{12} = 1 - 4 \mathcal{P}_{\bar{B^0} B^0}(\Delta t) + 4 Re[g(\Delta t)], 
\end{equation}
with 
\begin{equation}
g(t_1, t_2) = 2 \mathcal{P}_{\bar{B^0} B^0}(\Delta t) \mathcal{P}_{\bar{B}^0 \bar{B}^0}(\Delta t) + |\frac{p}{q}|^2 \frac{e^{-2 \Gamma \Delta t}(e^{-2 \lambda \Delta t} - 1)}{4} .
\end{equation}
Here $\mathcal{P}_{\bar{B}^0 \bar{B}^0}(\Delta t)$  and $\mathcal{P}_{\bar{B^0} B^0}(\Delta t)$  are the survival and transition probabilities, respectively, for the meson which  started in state $\rho_{\bar{B}^0} = \ket{\bar{B}^0}\bra{\bar{B}^0}$ at time $t=0$. The survival probability of $\bar{B}^o$ has the following form:
\begin{equation}
\mathcal{P}_{\bar{B}^0 \bar{B}^0}(t) = \frac{e^{-\Gamma t}}{2} \bigg[ \cosh(\frac{\Delta \Gamma t}{2}) + e^{-\lambda t} \cos(\Delta m t)\bigg] \label{Transition}.
\end{equation}
The LG function finally becomes
\begin{equation}
K_3 = 1 - 4 \mathit{P}_{\bar{B}^0 B^0}( \Delta t) + 8 \mathit{P}_{\bar{B}^0  B^0}( \Delta t) \mathit{P}_{\bar{B}^0  \bar{B}^0}( \Delta t) +  |p/q|^2 e^{-2 \Gamma  \Delta t} \big( e^{-2\lambda  \Delta t} - 1 \big). \label{K3}
\end{equation}
$CP$ violation implies that $|p/q| \ne 1$.
The above developed formalism also applies to the $K$ meson case with some notational changes. The $CP$ violating parameter for $K$ mesons $\epsilon$ can be expressed in terms of $p$ and $q$  by the following relation $\epsilon = \frac{p-q}{p+q} \label{epsilon}$.

%%%%%%%%%%%%%%%%%%%%%%%%%%%%%%%%%%%%%%%%%%
\subsection{Leggett-Garg type inequality}
%%%%%%%%%%%%%%%%%%%%%%%%%%%%%%%%%%%%%%%%%%%
The assumption of noninvasive measurability makes it difficult to test the Leggett-Garg inequality experimentally. Different measurement strategies like negative outcome measurement, delayed choice measurement, weak measurements \cite{tesche1990can, paz1993proposed, palacios2010experimental, goggin2011violation, fedrizzi2011hardy} have been devoted to this effect. Another formalism developed in \cite{huelga1995proposed, huelga1996temporal}, replacing the assumption of noninvasive measurability by ``stationarity", leads to easily testable inequalities using projective (von Neumann) measurements. According to the stationarity assumption, the conditional probability $q(t_i, t_j)$ to find a system in state $j$ at time $t_j$, if it was in state $i$ at time $t_i$ only depends on the time difference $t_j - t_i$, and is expected to hold not only for idealized closed quantum systems, but also in open quantum systems subjected to purely Markovian noise at a rate $\gamma$ such that the two time correlations are exponentially damped by a factor $\gamma (t_2 - t_1)$  \cite{waldherr2011violation}. The full set of assumptions (i)-(iv), for the stationarity to hold for a system, as given in Sec. (\ref{intro}), turns out to be applicable in the context of K and B meson systems. Given that the state of the meson at time $t=0$ is $\ket{\bar{B}^o}$, it can be shown that Markovian dynamics described by the Kraus operators in Sec. (\ref{dynamics})  lead to the time translation invariance of the conditional probability, i.e., $P(\bar{B}^o, t+t_0|\bar{B}^o,t_0) = P(\bar{B}^o,t|\bar{B}^o,0)$. With the assumption of stationarity, the Leggett-Garg type inequality, Eq. (\ref{tildeK3defined}), becomes
\begin{align}
\tilde{K}_3 &= 1 - 4 \mathit{P}_{\bar{B}^o B^o}( \Delta t) + 2 \mathit{P}_{\bar{B}^o B^o}(2 \Delta t). \label{lgtype}
\end{align}
	  Therefore, a knowledge of the transition probabilities at times $\Delta t$ and $2\Delta t$ would allow one to compute $K_3$ according to Eq. (\ref{lgtype}), such that $\tilde{K}_3 > 1$ shows the nonclassical nature of the neutral meson oscillations. It should be noted that Eq. (\ref{lgtype}) is expressed completely in terms of directly measurable quantities such as transition probabilities unlike Eq. (\ref{K3}), which contains a  term ($|p/q|^2 e^{-2 \Gamma  \Delta t} \big( e^{-2\lambda  \Delta t} - 1 \big)$), apart from the survival and transition probabilities.   However, it can be seen that in the limit of neglecting decoherence effects, Eq. (\ref{K3}), can also be expressed directly in terms of survival and transition probabilities. 
	
	\begin{figure*}[ht] 
	\centering
	\begin{tabular}{ccc}
		\includegraphics[width=60mm]{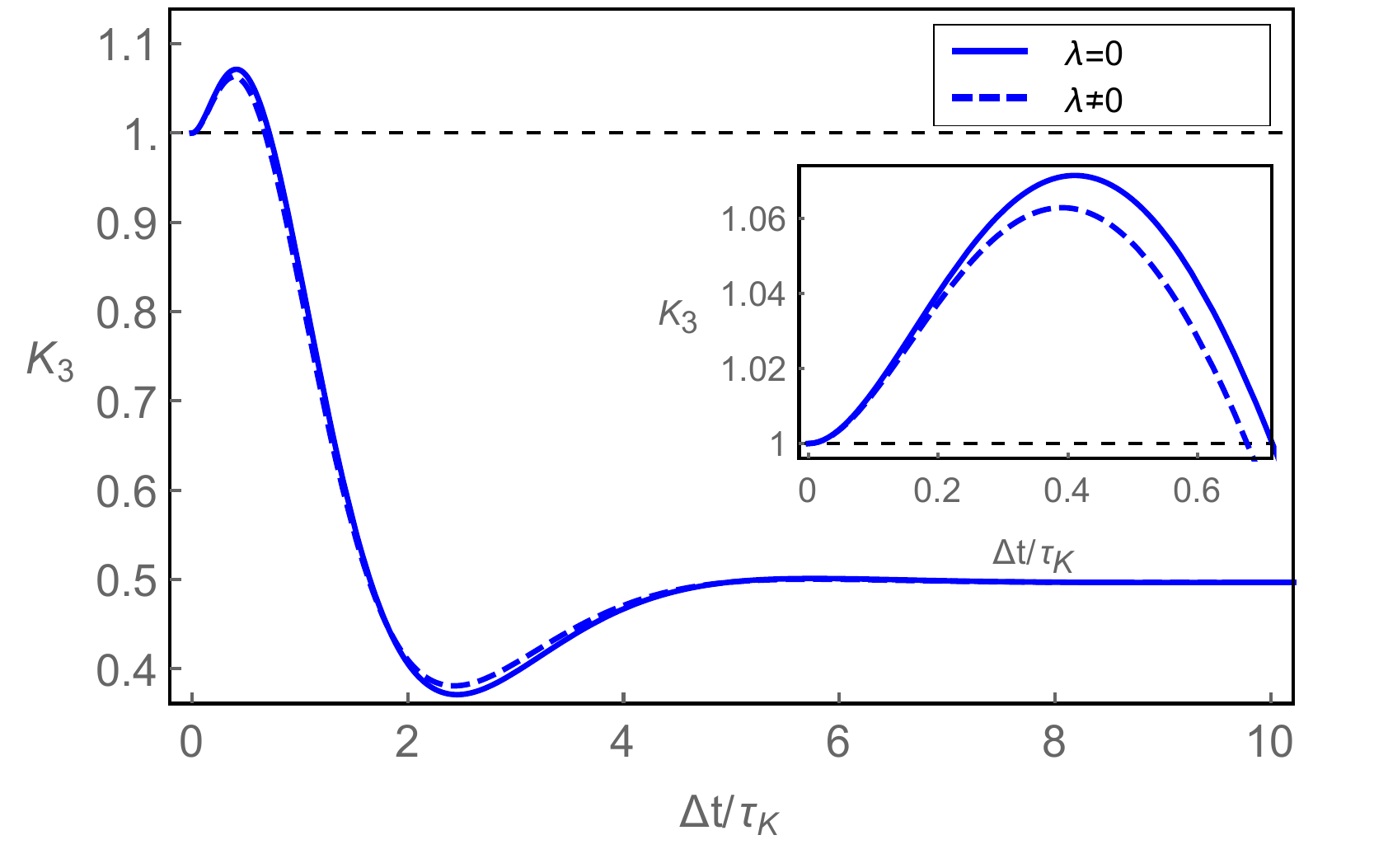} &
		\includegraphics[width=60mm]{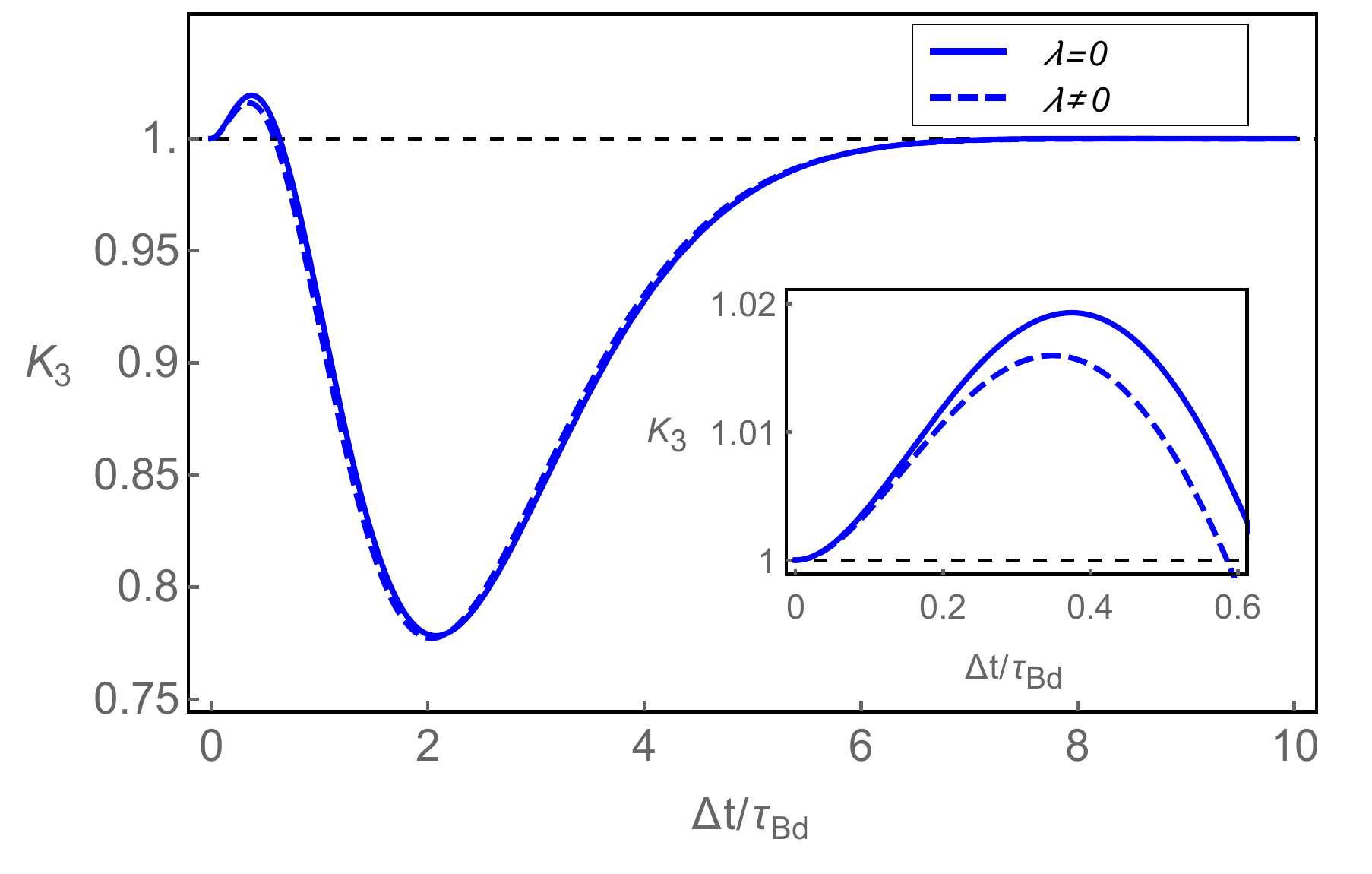}& 
		\includegraphics[width=60mm]{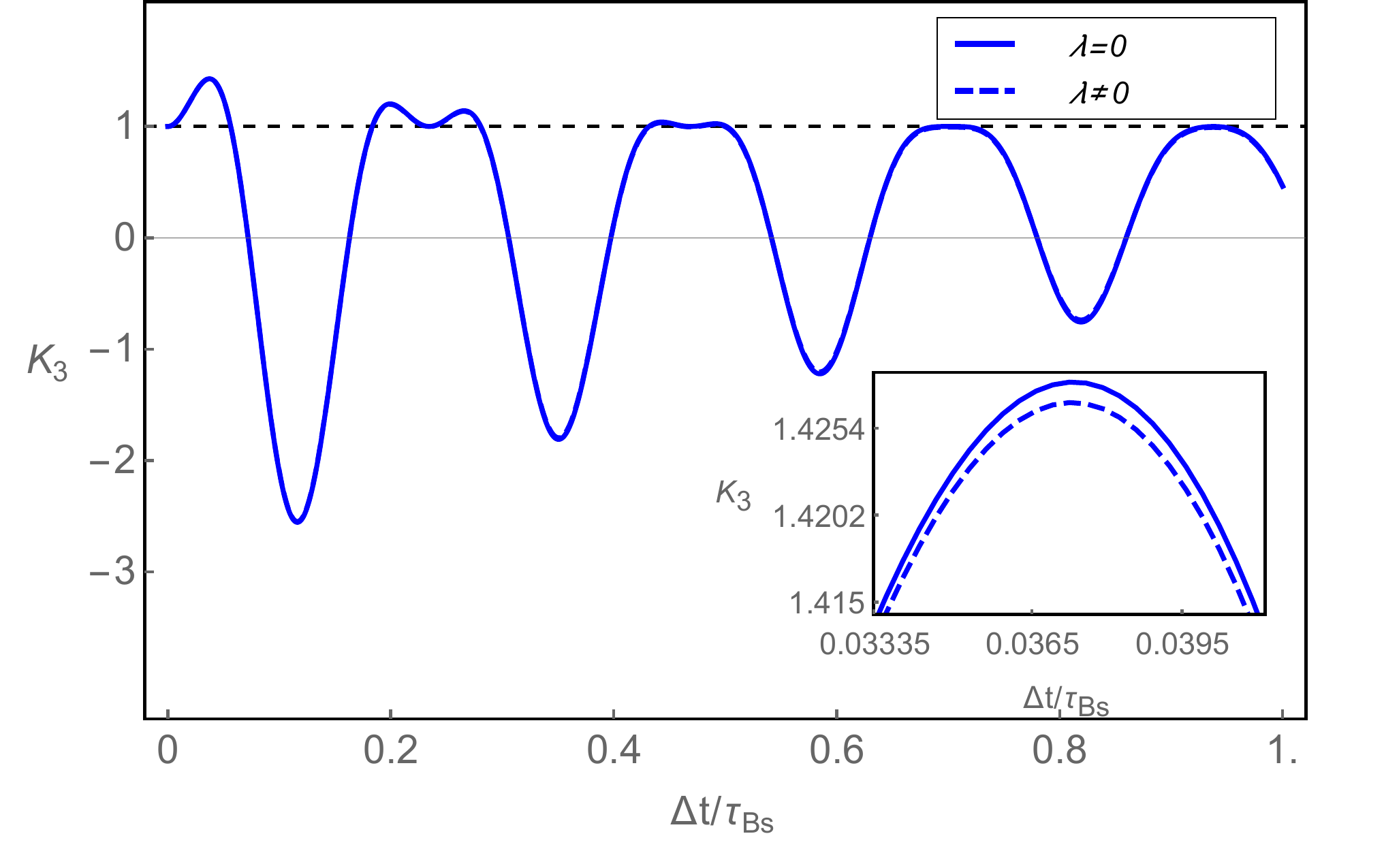} 
	\end{tabular}
	\caption{The left, middle and right panels of the figure depict the LG function $K_3$ plotted {\it w.r.t} the dimensionless quantity $\Delta t/\uptau$ for the $K$, $B_d$ and $B_s$ mesons, respectively.  Here $\Delta t$ is the time between successive measurements and $\tau$ is the mean lifetime of respective mesons. Dashed and solid curves correspond to the cases with and without decoherence, respectively. For the $K$ system, the mean lifetime is $  \uptau_K =  1.7889 \times 10^{-10} s $.  Also,  $\Gamma = 5.59 \times 10^{9}~ {\rm s^{-1}}$,  $ \Delta \Gamma = 1.1174 \times 10^{10}~{\rm s^{-1}}$,  $\lambda = 2.0 \times 10^{8}~ {\rm s^{-1}}$ and $\Delta m = 5.302\times 10^{9} ~ {\rm s^{-1}}$ \cite{Olive:2016xmw}. Here we used $ Re(\epsilon) = 1.596 \times 10^{-3} $ and $ |\epsilon| = 2.228 \times 10^{-3} $ \cite{d2006determination}. For the $B_d$ system, $\uptau_{B_d} = 1.518 \times 10^{-12} s $, $ \Gamma = 6.58 \times 10^{11}~ {\rm s^{-1}}$, $\Delta \Gamma = 0$,  $\lambda = 0.012 \times 10^{12} ~{\rm s^{-1}}$  and $\Delta m = 0.5064\times 10^{12} ~{\rm s^{-1}}$ \cite{Amhis:2016xyh}. The $CP$ violating parameter used here is $|\frac{q}{p}| = 1.010$ \cite{Amhis:2016xyh}.  Finally, for the $B_s$ meson, $\uptau_{B_s} = 1.509 \times 10^{-12} s $, $ \Gamma = 0.6645 \times 10^{12}~ {\rm s^{-1}}$, $\Delta \Gamma = 0.086 \times 10^{12}~ {\rm s^{-1}}$, $\lambda = 0.012 \times 10^{12}~ {\rm s^{-1}}$ and $\Delta m = 17.757\times 10^{12}~ {\rm s^{-1}}$ \cite{Amhis:2016xyh}. The value of the $CP$ violating parameter here is $\frac{q}{p} = 1.003$ \cite{Amhis:2016xyh}.  As we do not have any experimental bound on the decoherence parameter $\lambda$ for the $B_s$ system, we assume it to be the same as that of the $B_d$ system.}
\label{LG-meson}
\end{figure*}

 The experiments on the  $B^0(K^0)$ meson systems involve determination of their flavor at the time of production or decay. This is done by analyzing the flavor specific decays. For e.g., a $B^0_d$ meson can decay into a positron (or a $\mu^+$), a neutrino and a hadron with a branching ratio of $\sim 0.1$. This semileptonic decay is induced by the quark level transition $\bar{b} \to \bar{c}\, l^+ \,\nu_l $, with $l=e,\,\mu$.  On the other hand, the corresponding decay of a  $\bar{B^0_d}$ meson results in an electron (or a $\mu^-$) in the final state. Thus, in general, the charge of the final state lepton is same as the charge of the decaying quark. This is known as the $\Delta B =\Delta Q$ rule for the semileptonic decays of B mesons and is assumed in most of the experimental analysis. Hence, the charge of the final state lepton in the semi-leptonic decays of a neutral meson usually determines the flavor of that meson at the time of decay.

 The process of determination of the initial flavor of a neutral meson is called tagging. This is achieved by making use of the rule of associated production. The mesons are produced either by  strong or  electromagnetic interactions and hence  a quark is always produced in association with its anti-quark as flavor is conserved in these interactions. Thus, if a quark $q$ is detected at one end of the detector then at the quark at the other end has to be $\bar{q}$. Now if a charged meson is produced in association with a neutral meson, then the decay of the charged meson determines the flavor of the neutral meson at production. This is so because the charged meson cannot oscillate. The survival and oscillation probability of the neutral meson can then be measured by identifying the charge of the lepton in its semileptonic decay. If two entangled neutral mesons are produced,  as in the $e^+e^-$ colliders by the process $e^+e^- \to \Upsilon(4S) \to B^0_d \bar{B^0_d}$,  then detecting the flavor specific final state of one meson, say at time $t_1$, determines the flavor of that meson as well as the other meson at that time $t_1$. The oscillation probability of the tagged meson is then determined by identifying its final flavor specific state.
%%%%%%%%%%%%%%%%%%%%%%%%%%%%%%%%%%%%%%%%
\section{Results and discussion}
%%%%%%%%%%%%%%%%%%%%%%%%%%%%%%%%%%%%%%%

The left panel of Fig. (\ref{LG-meson}) shows the variation of the LG function $K_3$, as a function of the dimensionless quantity $\Delta t/\uptau_K$.  It can be seen from the figure that the LG  inequality is violated for about $\Delta t = \uptau_K $.  
The middle and right panels of Fig. (\ref{LG-meson}) depict the  variation of the LG function for the $B_d$ and $B_s$ mesons, respectively.  One can see that the violation in the $B_d$ meson system sustains for about $\Delta t = \uptau_{B_d}$  while for the $B_s$ meson system 
the violation is roughly for $\Delta t \approx 0.5~ \uptau_{B_s}$. The maximum violation of LGI occurs around $\Delta t \approx 0.41 \uptau_{K} $, $\Delta t \approx 0.37 \uptau_{B_{d}}$ and $\Delta t \approx 0.037 \uptau_{B_{s}}$ for $K$, $B_d$ and $B_s$  meson systems, respectively.  

 The figures clearly bring out the point that from the genesis of its decay \cite{Alok:2013sca}, the meson systems violate the upper threshold value of $K_3 = 1$, indicative of quantum behavior, and quickly fall below one. The  $K_3$ value for $K$ meson remains above one longest while $B_s$ does it for the shortest time. In addition, the $B_s$ meson exhibits an additional recurrence behavior. In order to have an understanding of this recurrence behavior, we re-write Eq. (\ref{K3}) as 
 \begin{widetext}
 \begin{align}
 K_3 &= 1 + |p/q|^2\bigg[2 e^{-(\Gamma + \lambda) \Delta t} \cos(\Delta m \Delta t) - e^{-2(\Gamma + \lambda) \Delta t} \cos(2\Delta m \Delta t) \nonumber - 2e^{-\Gamma \Delta t} \cosh(\Delta \Gamma \Delta t/2) + e^{-2\Gamma \Delta t} \cosh(\Delta \Gamma \Delta t)\bigg]. \label{K3_FullForm} \nonumber \\
 \end{align}
\end{widetext}
One can then see that the oscillating behavior in the case of $B_s$ meson system could be attributed to the mass term $\Delta m$ (Eq. (\ref{K3_FullForm})), which plays the role of frequency, and is more than 35 times the corresponding value for the $B_d$ meson system.

\begin{figure}[ht] 
\centering
\begin{tabular}{cc}
\includegraphics[width=70mm]{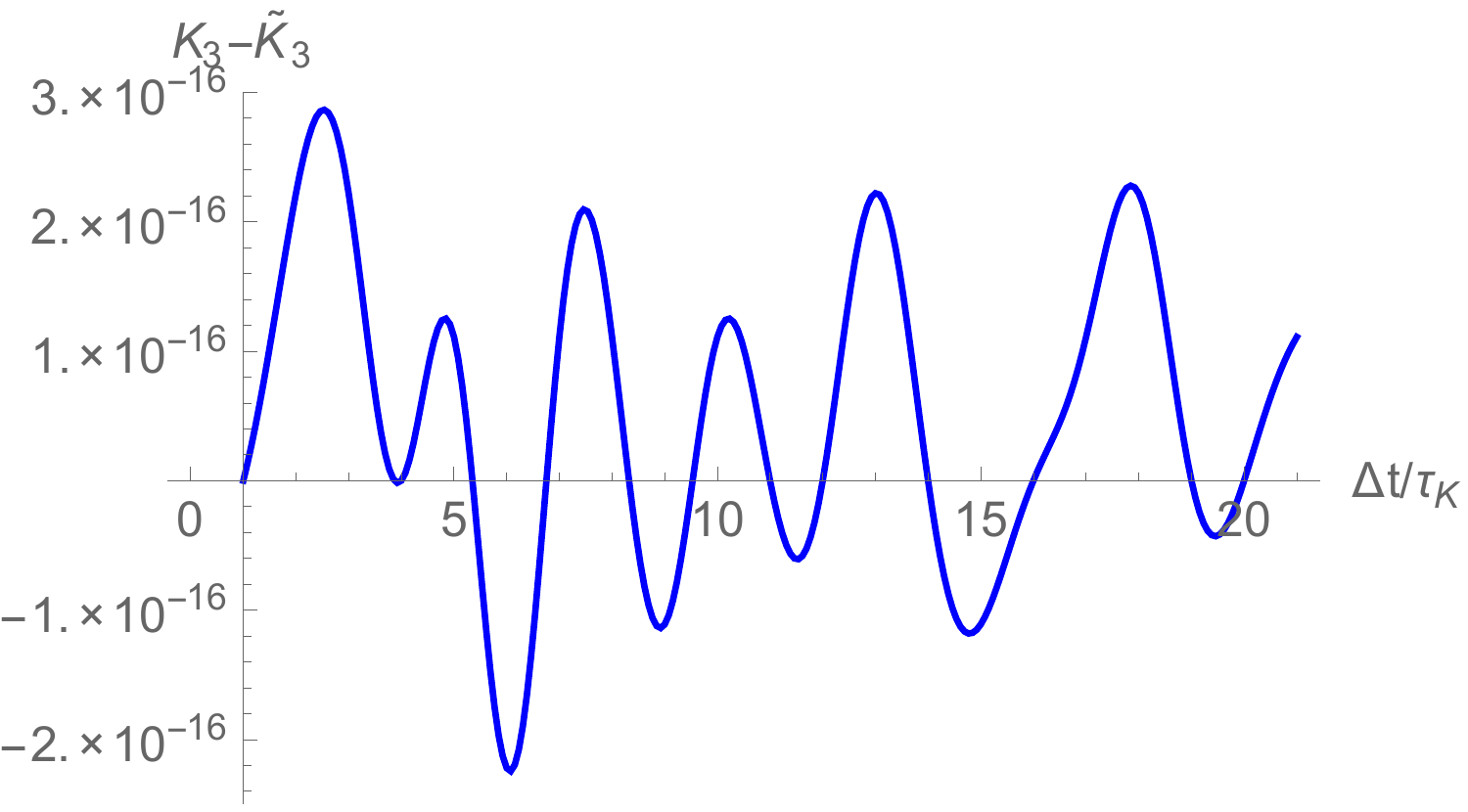}
\end{tabular}
\caption{ Plot of the difference of LG-function $K_3$ and LG-type function $\tilde{K_3}$ in the case of K meson system. The various parameters used are the same as in Fig. (\ref{LG-meson}).}
\label{K3minusK3}
\end{figure}

From Eq. (\ref{lgtype}), we find that the LG-type inequality is in terms of the transition probabilities only. Fig. (\ref{K3minusK3}) shows the deviation of the LG-type function, $\tilde{K_3}$ (\ref{lgtype}), from the LG-function ($K_3$). It is clear from the figure that the deviation is very small. Thus, a study of the LG inequality in mesons, using $\tilde{K_3}$, Eq. (\ref{lgtype}), in terms of experimentally measurable quantities would be well justified.   Eq. (\ref{lgtype}) demands the knowledge of the transition probabilities at $\Delta t$ and $2 \Delta t$, for example, ($0.5\uptau_{K}$, $\uptau_{K}$) for the $K$ meson system.

Looking at the form of Eq. (\ref{K3}), it can be seen that the only nonmeasurable term in the equation is $|p/q|^2 e^{-2 \Gamma  \Delta t} \big( e^{-2\lambda  \Delta t} - 1 \big)$; we call this term $\mathcal{D}_B$ and $\mathcal{D}_K$ for the case of $B$ meson and $K$ meson systems, respectively.  In the limit of zero decoherence, $\lambda \rightarrow 0$,  $\mathcal{D}_{B/K} \rightarrow 0$, rendering the LG function, Eq. (\ref{K3}), in terms of measurable survival and transition probabilities
\begin{align}
K_3(\lambda = 0) &= 1 - 4 \mathit{P}_{\bar{B}^0 B^0}( \Delta t) + 8 \mathit{P}_{\bar{B}^0  B^0}( \Delta t) \mathit{P}_{\bar{B}^0  \bar{B}^0}( \Delta t). \label{K3_lambda_zero}
\end{align}

\begin{figure*}[h] 
	\centering
	\begin{tabular}{ccc}
		\includegraphics[width=60mm]{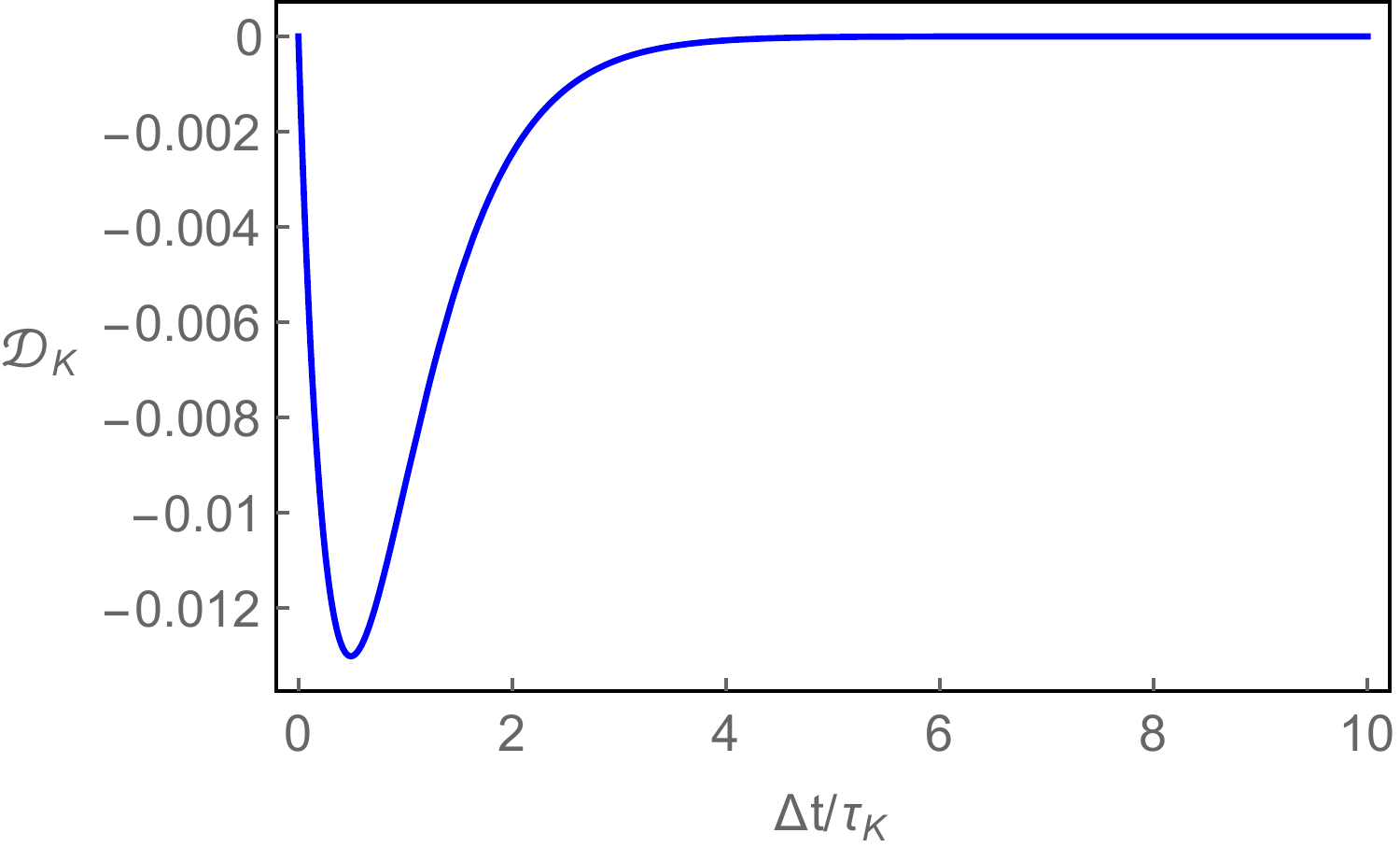}&
		\includegraphics[width=60mm]{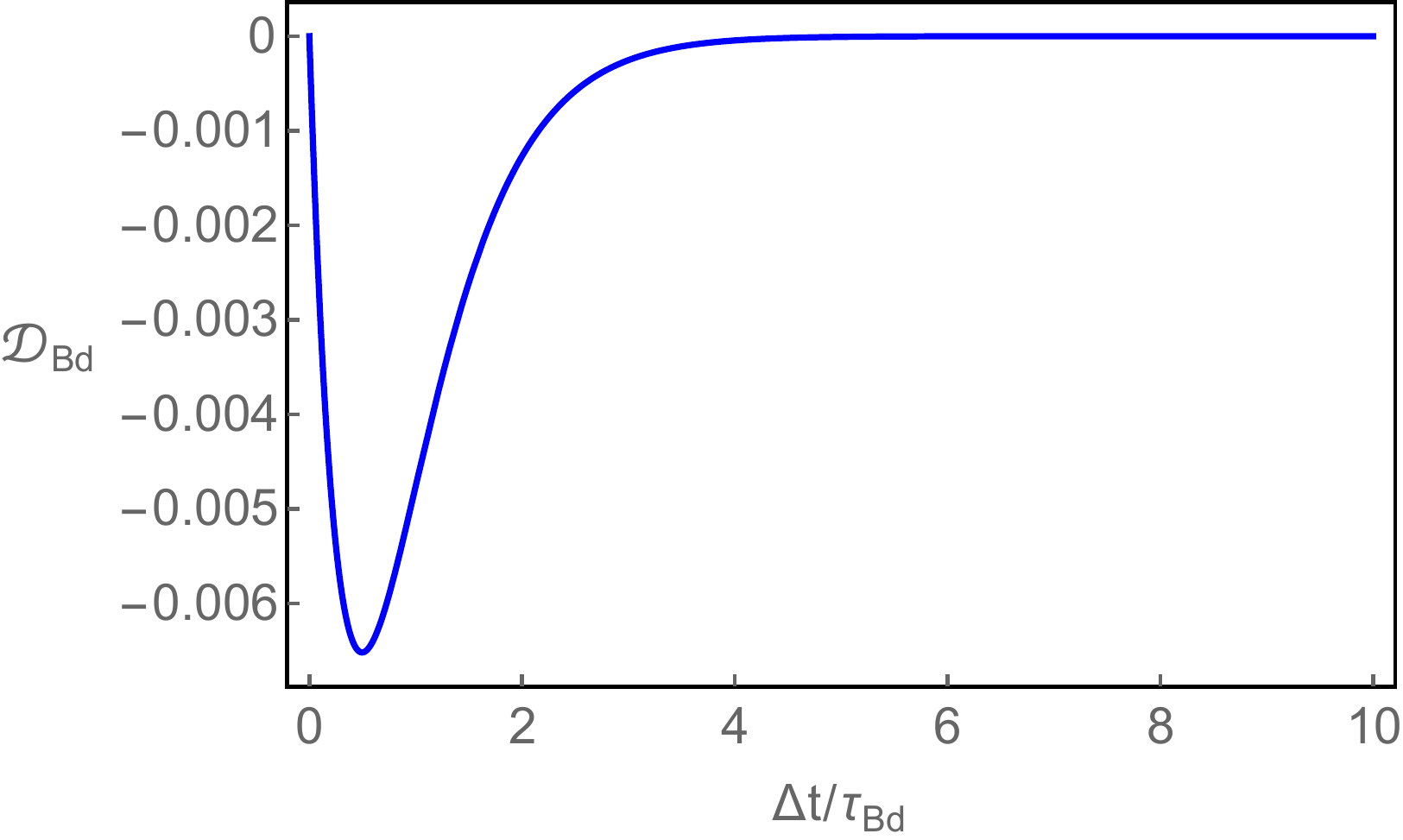}&
		\includegraphics[width=60mm]{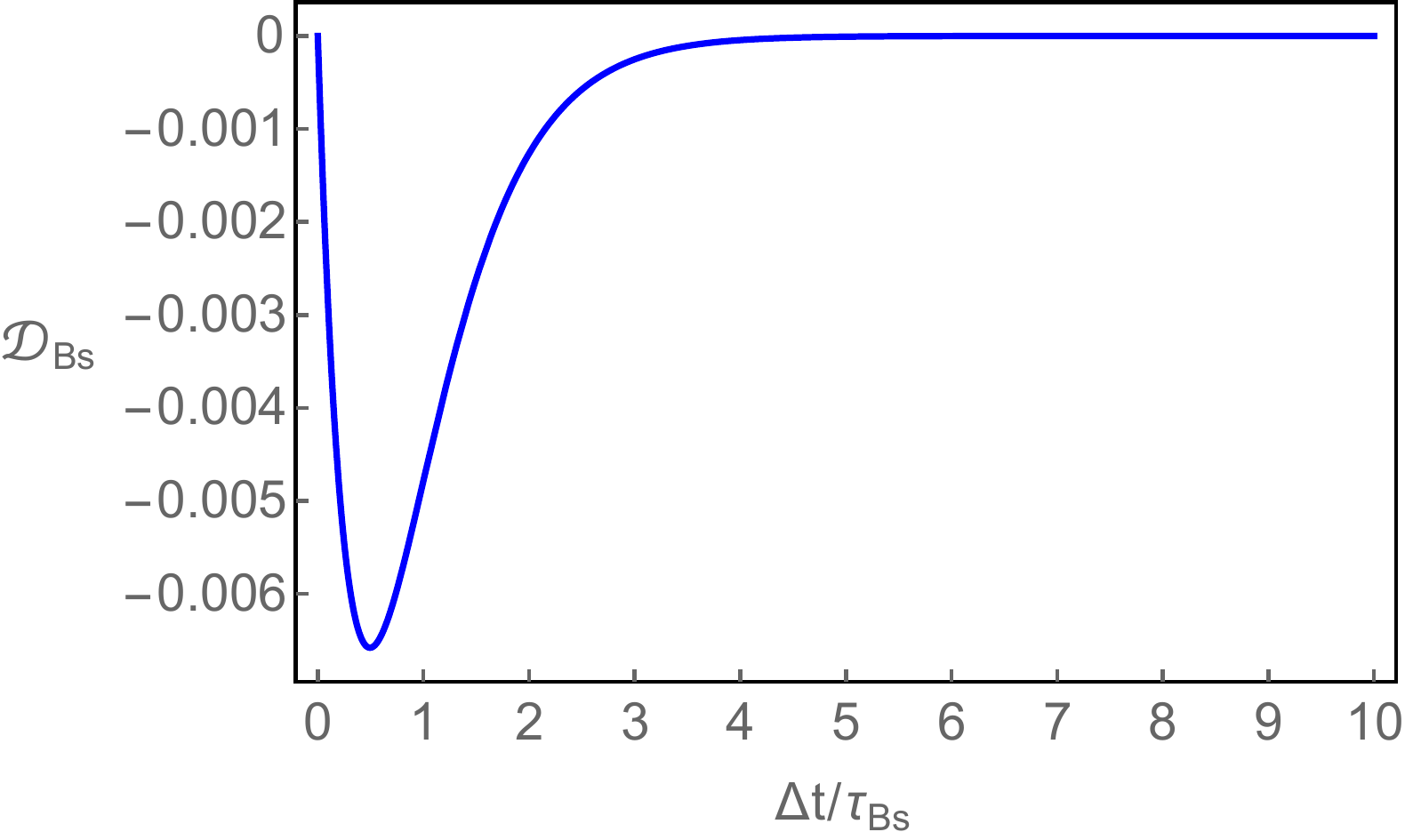}
	\end{tabular}
	\caption{ The nonmeasurable term $ \mathcal{D}_K = |(1+\epsilon)/(1-\epsilon)|^2 e^{-2 \Gamma  \Delta t} \big( e^{-2\lambda  \Delta t} - 1 \big)$ for $K$-meson system  and  $ \mathcal{D}_{Bd(s)} = |p/q|^2 e^{-2 \Gamma  \Delta t} \big( e^{-2\lambda  \Delta t} - 1 \big)$ for $B_{d(s)}$-meson system, plotted against $\Delta t/\uptau_{K/B_{d(s)}}$. The various parameters used in the two cases are the same as mentioned in the caption of Fig.~(\ref{LG-meson}).}
	\label{DKDM}
\end{figure*}

 The variation of $\mathcal{D}_B$ and $\mathcal{D}_K$  with $\Delta t/\uptau_{K/B_{d(s)}}$ is shown in Fig.~\ref{DKDM}. 
It is obvious from the figure that these terms are small compared with the maximum value attained by the LG function $K_3$. 

%%%%%%%%%%%%%%%%%%%%%%%%%%%%%%%%%%%
\section{Conclusion}
%%%%%%%%%%%%%%%%%%%%%%%%%%%%%%%%%%
In this work, we study the violation of LG and LG-type inequalities in $B$ and $K$ mesons within the framework of open quantum systems. It is found that LGI is violated in both $K$ and $B$ meson systems. This violation lasts for a longer time in the case of $K$ mesons as compared to that of $B$ mesons.  In the case of $B$ meson systems, the violation lasts longer for $B_d$ mesons as compared to the $B_s$ system.  We show that the LG function $K_3$, apart form the measurable survival and transition probabilities, contains a  nonmeasurable term which is small compared to the maximum value attained by it and vanishes in the approximation of zero decoherence.  Since systems with no coherence do not violate LGI, the effect of \textit{decoherence} should result in decreasing the extent of the violation, as observed in Fig. (\ref{LG-meson}).   Further, it is highlighted in this work that the LG-type function, unlike LG function, can be expressed completely in terms of experimentally measurable quantities.  Hence, LG type inequality is  seen to be more suitable for understanding the nature of temporal quantum correlations in meson systems.

%\bibliographystyle{apsrev4-1}
%\bibliography{References_MesonLGI}

%merlin.mbs apsrev4-1.bst 2010-07-25 4.21a (PWD, AO, DPC) hacked
%Control: key (0)
%Control: author (72) initials jnrlst
%Control: editor formatted (1) identically to author
%Control: production of article title (-1) disabled
%Control: page (0) single
%Control: year (1) truncated
%Control: production of eprint (0) enabled
%

\end{document}